\documentclass[12pt]{iopart}
\pdfoutput=1
\usepackage{graphicx}
\usepackage{wrapfig}
\expandafter\let\csname equation*\endcsname\relax
\expandafter\let\csname endequation*\endcsname\relax
\usepackage{amsmath}
\usepackage{color}
\usepackage{url}
\usepackage{dcolumn}
\usepackage{bm}
\newcommand{\eqb}{\begin{eqnarray}}
\newcommand{\eqe}{\end{eqnarray}}
\newcommand{\diff}{\textrm{d}}

\newcommand{\lambdamicron}{\lambda_{\mu\textrm{m}}}

\newcommand{\wsqcm}{\textrm{W\,cm}^{-2}}

\newcommand{\uperp}{{\bm{u}_\bot}}
\newcommand{\pperp}{{\bm{P}_\bot}}
\newcommand{\pr}{P_r}
\newcommand{\pphi}{\Theta}
\newcommand{\sign}{s}
\newcommand{\etahat}{\hat{\eta}}
\newcommand{\prhat}{\hat{P}_r}
\newcommand{\rr}{r_{\textrm{c}}}
\newcommand{\rrmax}{r_{\textrm{max}}}
\newcommand{\rrmin}{r_{\textrm{min}}}

\newcommand{\dynamicalsystem}{(\ref{xdoteq}--\ref{pphidoteq})}

\begin{document}
\title{Radiative trapping in intense laser beams}

\author{J G Kirk}

\address{Max-Planck-Institut f\"ur Kernphysik, 
Postfach 10 39 80, 69029 Heidelberg, Germany}
\ead{john.kirk@mpi-hd.mpg.de}
\begin{abstract}
  The dynamics of electrons in counter-propagating, circularly
  polarized laser beams are shown to exhibit attractors whose ability
  to trap particles depends on the ratio of the beam intensities and a
  single parameter describing radiation reaction. Analytical
  expressions are found for the underlying limit cycles and the
  parameter range in which they are stable. In high-intensity optical 
  pulses, where radiation reaction strongly modifies the trajectories,
  the production of collimated gamma-rays and the initiation of
  non-linear cascades of electron-positron pairs can be optimized by a
  suitable choice of the intensity ratio.
\end{abstract}

\pacs{12.20.-m, 52.27.Ep, 52.38.Ph}
Journal reference: {\em  Plasma Physics and Controlled Fusion\ {\bf 8}, 085005 (2016)}

\section{Introduction}
The design of experiments on 10~PW laser facilities that will probe
strong-field QED
\cite{ELI,XCELS,HiPER} requires an
understanding of the behaviour of electrons and positrons in intense
photon beams \cite{dipiazzaetal12}. However, in all but the simplest
configurations, such as a single plane wave \cite{dipiazza08}, the
dynamics are highly complex, displaying regions of both ordered and
chaotic motion
\cite{mendoncadoveil82,bauermulsersteeb95,lehmannspatschek12}.
Radiative reaction plays a crucial role at these intensities, and has
been implemented in numerical studies by several groups, using both a
classical and a fully quantum approach
\cite{zhidkovetal02,naumovaetal09,tamburinietal10,chenetal11,nakamuraetal12,ridgersetal14,bulanovetal15}.
Amongst other effects, it leads to a contraction of phase-space, which
can cause particle trajectories to accumulate on attractors
\cite{lehmannspatschek12,gonoskovetal14,jietal14,esirkepovetal15,jirkaetal16}.
This \lq\lq radiative trapping\rq\rq\ is an important
effect, because it may allow the electrons and the
gamma-rays they radiate to be controlled.

Attractors are closely related to limit cycles, which are 
periodic trajectories in phase space. When these are stable
to infinitesimal perturbations, they correspond to a
simple attractor, which confines 
trajectories that pass sufficiently close to the cycle. 
When they are unstable, an attractor with a more complicated structure 
(a strange attractor) may
nevertheless be present in the immediate neighbourhood, and again lead
to a concentration of particles in a restricted region of phase-space.
The goal of this paper is to identify electron limit cycles
analytically and to study their stability properties using both
analytical and numerical tools. To achieve this, the treatment is
restricted to classical electrodynamics. The chosen field
configuration consists of two circularly polarized, monochromatic,
vacuum waves with arbitrary frequencies, amplitudes and propagation
directions, but with opposite helicities.  In classical
electrodynamics, only two dimensionless parameters are needed to
specify the relativistic dynamics in this set-up. These are
the ratio $\lambda$ of the wave amplitudes, and the
classical radiation reaction parameter $\rr$, defined in
equation~(\ref{rrdef}).  To date, work in this field has concentrated on
numerical integration of trajectories in a standing wave, which
corresponds to $\lambda=1$. In this case, circularly polarized waves
exhibit an attractor only where the electric field vanishes, and
strong-field QED effects, such as the production of hard gamma-rays and
electron positron pairs, are negligible.

In this paper, attractors are found on which the electric field
amplitude and the particle energy are sufficient to permit
strong-field QED effects in a limited range of $\lambda$.  Expressions
giving this range, the limit-cycle orbits and their linear stability
properties as functions of $\lambda$ and $\rr$ are presented in
section~\ref{analyticalresults}. Following this, numerical integration
of the equations of motion is used in section~\ref{numericalresults}
to present stroboscopic sections of phase-space illustrating the
character of the particle orbits close to these limit cycles.  The
remainder of the paper is organized as follows:
section~\ref{equations} introduces the notation, presents the field
configuration and formulates the equations of motion,
section~\ref{discussion} relates the results to previous work and
briefly discusses some practical implications, and
section~\ref{conclusions} summarizes the conclusions.

\section{Field configuration and equations of motion}
\label{equations}
Consider two monochromatic plane
waves in vacuum, with wave vectors $\vec{k}_{1,2}/c$ and frequencies
$\omega_{1,2}=\left|\vec{k}_{1,2}\right|$. Assuming
$\vec{k}_1\ne\vec{k}_2$, it is always possible to find a frame of
reference in which these waves have the same frequency and oppositely
directed wave vectors. In a system consisting of a single photon from
wave~1, and a single photon from wave~2, this is the frame in which
the total momentum is zero (the ZMF). In it, the waves have the frequency
$\omega=\left(\omega_1\omega_2-\vec{k}_1\cdot\vec{k}_2\right)^{1/2}/\sqrt{2}$.

Our strategy is to look for limit cycles that lie entirely in the
transverse plane, when viewed in the ZMF. On such orbits, the force
exerted by the magnetic field must vanish. One obvious location is a
plane in which the magnetic field itself vanishes, but this can occur 
only in very special configurations, such 
counter-propagating waves of equal amplitude and either aligned linear
polarization or circular polarization of opposite helicity. More
generally, one can search for orbits on which the particle velocity is
always parallel to the magnetic field. Such orbits occur when the
electric and magnetic field vectors corotate in the transverse plane
such that their phase difference and amplitudes remain constant.  In
the remainder of this section, we show that this configuration is
present in counter-propagating waves of arbitrary relative amplitude,
provided they are of opposite helicity, and formulate the
corresponding equations of motion.

The electric field is related to the dimensionless vector potential
$\vec{a}$ by
$\vec{E}_{1,2}=\left(mc/e\right)\partial\vec{a}_{1,2}/\partial t$,
where $m$ and $e$ are the mass and charge of the positron.  For
circularly polarized waves, one can choose a Lorentz gauge in which
$\left|\vec{a}\right|=\textrm{constant}$ and the electrostatic
potential vanishes.  The ratio of the magnitudes of the dimensionless
vector potentials can be defined as $\lambda=a_2/a_1$, where, without
loss of generality, $0\le\lambda\le1$.  Note that $a_{1,2}$ and
$\lambda$ are Lorentz invariant, as is also the helicity of each
wave. Thus, in terms of dimensionless independent variables
($t\rightarrow t/\omega$, $x\rightarrow xc/k$), the transverse
components of the vector potentials in the ZMF can be written as
\begin{align}
\bm{a}_1&=\frac{a_0}{2}\textrm{e}^{\pm\imath \left(t-x+\phi_1\right)}
\nonumber\\
\bm{a}_2&=\frac{\lambda a_0}{2}\textrm{e}^{\mp\imath \left(t+x+\phi_2\right)}\,.
\label{definitiona0}
\end{align}
(Here, and in the following, the complex notation $\bm{a}\equiv
\vec{a}\cdot\left(\hat{\vec{y}}+\imath \hat{\vec{z}}\right)$ is used
for the transverse components of a vector.) 
In (\ref{definitiona0}) the upper (lower) signs correspond to left-(right-)handed
polarizations, the $\phi_{1,2}$ are arbitrary phases, and the $x$-axis is chosen to lie in the direction of propagation of the 
primary wave (wave~1). The constant $a_0$, defined in equation (\ref{definitiona0}) is 
the magnitude of the vector potential in the standing wave ($\lambda=1$), and will be used below to normalize 
the particle four-momentum. At fixed $x$, the
amplitude of the superposed fields is constant in time, provided waves
of opposite helicity are selected.  Selecting convenient phases, and
choosing left-handed polarization for
wave~1, the vector potential is
\begin{align}
\bm{a}&=\bm{a}_1+\bm{a}_2
\nonumber\\
&=\imath a_0\textrm{e}^{\imath t}\left(
\textrm{e}^{-\imath x}+\lambda\textrm{e}^{\imath x}\right)/2
\label{superposedpotential}\,.
\end{align}
Thus, in addition to the amplitudes, also the relative phase of the 
$\vec{E}$ and $\vec{B}$ vectors  
at a given location is constant in time:
\begin{align}
\left|\vec{E}\right|&= \left(\frac{mc\omega}{e}\right)\frac{a_0}{2}\left[
1+\lambda^2+2\lambda\cos\left(2x\right)\right]^{1/2}
\nonumber\\
\left|\vec{B}\right|&= \left(\frac{mc\omega}{e}\right)\frac{a_0}{2}\left[
1+\lambda^2-2\lambda\cos\left(2x\right)\right]^{1/2}
\nonumber\\
\vec{E}\cdot\vec{B}&=
\left(\frac{mc\omega}{e}\right)^2 a_0^2\lambda\sin x\cos x\,.
\label{edotb}
\end{align}
We now set $m=c=1$ and write the (dimensionless) four-momentum, as $(\gamma,u_x,\uperp)$, where
$\uperp=u_y+\imath u_z$ 
and $\gamma=\left(1+u_x^2+\uperp\uperp^*\right)^{1/2}$, and further introduce the
canonical momentum in the transverse plane $\pperp$ and its radial and
angular components $\pr$ and $\pphi$ according to
\begin{align}
\pperp&\equiv\sign \pr\textrm{e}^{\imath\pphi}
\\
&=\uperp +\sign \bm{a}\,,
\end{align}
where $\sign=q/e$, with $q$ the particle charge. 
In the absence of radiation reaction, the dynamics are Hamiltonian, with 
the three canonical momenta $u_x$, $\pr$ and $\pphi$. The Hamiltonian function
\begin{align}
H(u_x,\pr,\pphi,x)&=\gamma 
\nonumber\\
&=\left\lbrace
1+u_x^2+\pr^2+\frac{a_0^2}{4}\left(1+\lambda^2+2\lambda\cos2x\right)
\right.
\nonumber\\
&\left.\phantom{\frac{0}{0}}
+a_0\pr\left[\sin\left(t-\pphi-x\right)+\lambda\sin\left(t-\pphi+x\right)\right]
\right\rbrace^{1/2}
\label{hamiltoniandef}
\end{align}
does not contain the coordinates conjugate to $\pr$ and $\pphi$,
indicating that these are constants of the motion. Ignoring the corresponding 
coordinates, the two remaining first-order equations of motion are
\begin{align}
\frac{\diff x}{\diff t}&=u_x/\gamma
\\
\frac{\diff u_x}{\diff t}&=\frac{1}{\gamma}\left[\lambda a_0^2\sin x\cos x+\frac{a_0\pr}{2}\cos\left(t-\pphi-x\right)
-\frac{\lambda a_0\pr}{2}\cos\left(t-\pphi+x\right)\right]\,.
\label{norrsys}
\end{align}
Interestingly, the choice
$\pr=0$ renders the Hamiltonian independent of time, so that the third
constant of motion is $\gamma$ and the system is integrable.  

To take account of radiation reaction, we adopt the approximation used
in \cite{kirkbellarka09}, in which the spatial components of the
four-force, are anti-parallel to those of the four-momentum.  This
approximation arises from the leading term in an expansion in
$1/\gamma$ of the radiation reaction force as formulated by
\cite{landaulifshitz75} and is adequate to describe the orbits in
large amplitude, circularly polarized waves. The radiation reaction
terms are then
\begin{align}
\left(\frac{\diff \vec{u}}{\diff t}\right)_{RR}&=
-\epsilon\left(\frac{mc^2}{\hbar\omega}\right)^2\eta^2\vec{u}/\gamma\,,
\label{rrterms}
\\
\noalign{\hbox{where}}
\epsilon&=\frac{2e^2\omega}{3mc^3}\ ,
\nonumber
\end{align}
and $\eta$ is the parameter that determines the importance of QED
effects, defined in terms of the electromagnetic field tensor, $F^{\mu\nu}$, as
\begin{align}
\eta&=\left|F^{\mu\nu}u_\nu\right|/E_{\textrm{crit}}\,,
\label{etadef}
\end{align}
with $E_{\textrm{crit}}=m^2c^3/e\hbar$ the critical or {\em Schwinger}
field. Thus, retaining only the leading terms in $1/\gamma$, the
equations of motion are
\begin{align}
\frac{\diff x}{\diff t}&=\hat{u}_x/\hat{\gamma}
\label{xdoteq}
\\
\frac{\diff \hat{u}_x}{\diff t}&=\frac{1}{2\hat{\gamma}}\left[\lambda \sin 2x +\prhat\cos\left(t-\pphi-x\right)
-\lambda \prhat\cos\left(t-\pphi+x\right)\right]
-\rr\frac{\etahat^2\hat{u}_x}{\hat{\gamma}}
\label{uxdoteq}
\\
\frac{\diff \prhat}{\diff t}&=-\rr\frac{\etahat^2}{2\hat{\gamma}}\left[2\prhat+\sin\left(t-\pphi-x\right)
+\lambda\sin\left(t-\pphi+x\right)\right]
\label{prdoteq}
\\
\frac{\diff \pphi}{\diff t}&=\rr
\frac{\etahat^2}{2\hat{\gamma}\prhat}\left[\cos\left(t-\pphi-x\right) + \lambda\cos\left(t-\pphi+x\right)\right]\,,
\label{pphidoteq}
\end{align}
where we have introduced the classical radiation reaction parameter
\begin{align}
\rr&=\epsilon a_0^3
\label{rrdef}
\end{align}
\cite{dipiazzaetal12,kirkbellridgers13}, and the 
four-momentum has been normalized to $a_0$:
\begin{align}
\hat{u}_x&=u_x/a_0
\\
\prhat&=\pr/a_0
\\
\hat{\gamma}&=\left\lbrace
\hat{u}_x^2+\prhat^2+\frac{\left(1+\lambda^2+2\lambda\cos2x\right)}{4}
+\prhat\left[\sin\left(t-\pphi-x\right)+\lambda\sin\left(t-\pphi+x\right)\right]
\right\rbrace^{1/2}\,.
\label{gammahatdef}
\end{align}
For the transverse wave-fields 
given by equation~(\ref{superposedpotential}), $\eta$ can be written
\begin{align}
\eta&=\left(\frac{\hbar\omega}{mc^2}\right)a_0^2\etahat\,,
\label{etahatdef2}
\\
\noalign{\hbox{where}}
\etahat&\approx
\frac{1}{2}\left\lbrace
\left(1+\lambda^2\right)
\left(\hat{\gamma}^2+\hat{u}_x^2\right)
+2\lambda\cos2x\left(\hat{\gamma}^2-\hat{u}_x^2\right)-2\left(1-\lambda^2\right)\hat{\gamma}
\hat{u}_x
\phantom{\left[2\prhat\right]}\right.
\nonumber\\
&\left.
-\lambda\left[2\prhat\cos\left(t-\pphi+x\right)-\sin2x\right]
\left[2\prhat\cos\left(t-\pphi-x\right)
+\lambda \sin2x\right]\right\rbrace^{1/2}\,.
\label{etahatdef}
\end{align}
(The approximation made in equation (\ref{etahatdef}) consists of
making the replacement $\gamma\rightarrow a_0\hat{\gamma}$.)  Therefore,
provided quantum effects are negligible and the motion is
relativistic, the equations of motion \dynamicalsystem\ depend on only
two parameters: $\lambda$ and $\rr$.

\section{Limit cycles and stability}
\label{analyticalresults}
Limit cycle solutions to the system \dynamicalsystem\ can be located
by searching for points at which the parallel momentum $\hat{u}_x$
vanishes (and, hence, $x$ is constant), the azimuthal momentum $\pphi$
is proportional to $t$, and the radial momenta $\prhat$ is constant.
Using the latter requirement, equation~(\ref{prdoteq}) determines $\prhat$.
Substituting this expression into Eq~(\ref{uxdoteq}), and setting $\hat{u}_x=0$
and $\pphi=t+\theta_0$, one finds that $\diff \hat{u}_x/\diff t$
vanishes when
\begin{align}
\theta_0&=
\textrm{arctan}\left(\frac{-\left(1+\lambda\right)\tan x}{\left(1-\lambda\right)}\right)\,,
\label{theta0fix}
\end{align}
at which point
\begin{align}
\prhat&=-\frac{2\lambda\hat{\gamma}}{1-\lambda^2}\sin\left(2x\right)
\label{prfix}
\\
\noalign{\hbox{and}}
\etahat&=\hat{\gamma}^2\,.
\label{gammafix}
\end{align}
Finally, the self-consistency requirement $\diff\pphi/\diff t=1$,
determines the values of the initial phase $\theta_0$. If it can be
fulfilled at a point $x$, limit cycles exist at $x+n\pi$ for integer
values of $n$, and consist of circular trajectories in the $y$-$z$
plane. The trajectories at even and odd values of $n$ are essentially
identical, i.e., particles rotate in the
same sense but with a phase difference of $\pi$.

\begin{figure}
\includegraphics{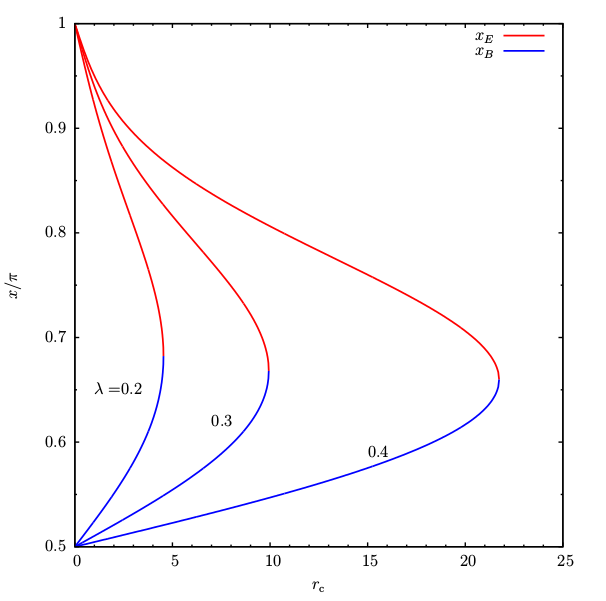}
\caption{\label{xpoints}
The location of the limit cycles as a function of the 
radiation reaction parameter $\rr$ [see equation~(\ref{rrdef})] for three values of the 
ratio $\lambda$ of the wave amplitudes. The suffix $E$ ($B$) denotes the cycle that 
is located close to the antinode of the electric (magnetic) field in the 
standing wave that exists when $\lambda=1$.
 }
\end{figure}

These points can be found by considering $\diff\pphi/\diff t$
as a function of $\rr$, $\lambda$ and $x$, using (\ref{prfix}) to
substitute for $\prhat$ and (\ref{theta0fix}) to substitute for $\theta_0$.
Then, for fixed $\lambda$, the function $\diff\pphi/\diff t-1$ has 
roots which coincide with those of a fifth order polynomial in $\tan^2x$.
For small $\rr$, there are two roots, which are conveniently chosen to 
lie in the range $\pi/2\le x\le\pi$. 
As
$\rr$ increases the roots approach each other and vanish for $\rr$
larger than a critical value, given by:
\begin{align}
\rrmax\left(\lambda\right)&=\frac{32\lambda}{\left(1-\lambda\right)^4}
\left[\frac{\left(8\lambda+\Delta\right)\left(1+6\lambda+\lambda^2+\Delta\right)^3}
{\left(1+10\lambda+\lambda^2+\Delta\right)^5}\right]^{1/2}\,,
\\
\noalign{\hbox{where}}
\Delta&=\sqrt{1+62\lambda^2+\lambda^4}\,.
\nonumber
\end{align}

For $\rr\ll \lambda$, the limit cycles lie close to $\pi/2$ and $\pi$, 
and approximate expressions can be found. For the cycle that is close to 
$x=\pi/2$ (which is an antinode of the magnetic field when $\lambda=1$,
hence the suffix $B$) one has, to lowest order in $\rr$ and for fixed $\lambda$:
\begin{align}
x&=x_B\,\approx\,\frac{\pi}{2}+\frac{\rr\left(1-\lambda\right)^4
\left(1+\lambda\right)}{32\lambda}
\label{xbapprox}
\\
{\prhat}&\,\approx\,\frac{\rr\left(1-\lambda\right)^4}{16}
\\
{\pphi}&\,\approx\, t-\frac{\pi}{2}-\frac{\rr\left(1-\lambda\right)^5}{32\lambda}
\\
\hat{\gamma}&\,\approx\,\frac{1-\lambda}{2}
\label{gammaapprox}
\\
\etahat&\,\approx\,\frac{\left(1-\lambda\right)^2}{4}
\label{etaapprox}
\end{align}
and for the cycle that is close to $\pi$ (an antinode of the electric field 
when $\lambda=1$):
\begin{align}
x&\,=x_E\,\approx\,\, \pi-\frac{\rr\left(1-\lambda\right)
\left(1+\lambda\right)^4}{32\lambda}
\\
{\prhat}&\,\approx\,\frac{\rr\left(1-\lambda\right)^4}{16}
\\
{\pphi}&\,\approx\, t+\pi+\rr\frac{\left(1+\lambda\right)^5}{32\lambda}
\\
\hat{\gamma}&\,\approx\,\frac{1+\lambda}{2}
\\
\etahat&\,\approx\,\frac{\left(1+\lambda\right)^2}{4}\,.
\end{align}
In the general case, a root-finding algorithm is necessary in order to locate 
$x_E$ and $x_B$; 
the results are shown in Fig~\ref{xpoints}.

\begin{figure}
\includegraphics{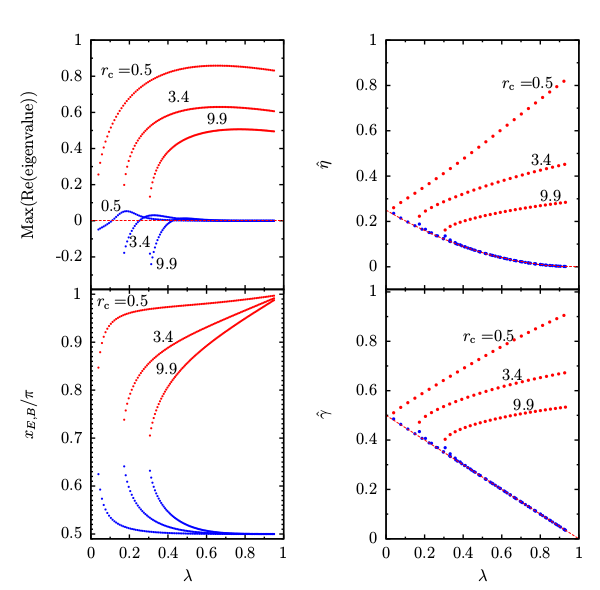}
\caption{\label{stationary_sols} Properties of the limit cycles for
  three values of $\rr$. The cycles lie in the $y$-$z$ plane, at the
  location shown in the lower left panel. The top left panel shows the
  largest of the real parts of the eigenvalues of the Jacobian (\ref{jacobian}). 
  Cycles located at $x_E$, plotted in red,
  are unstable for all $\rr$, those at $x_B$, shown in blue, have a
  region of stability below a critical value of $\lambda$, but are
  unstable above it.  The cycles consist of circular motion at
  constant Lorentz factor, and constant value of the QED parameter
  $\eta$. The scaled values of these quantities $\hat{\gamma}$ and
  $\etahat$ [see equations~(\ref{gammahatdef}) and (\ref{etahatdef2})] are
  shown in the lower right and upper right panels, respectively, which also plot the 
approximations given by equations~(\ref{gammaapprox}) and (\ref{etaapprox}) as dashed, red lines. }
\end{figure}

It is easy to understand on physical grounds that such a limit cycle
cannot exist for all values of $\lambda$ and $\rr$. In a single,
circularly polarized plane wave in the absence of radiation reaction,
a circular trajectory that lies in the $y$-$z$
plane can always be found by choosing a suitable value of
$\prhat$. However, the radiation reaction force acts to push the particle
in the direction of propagation of the wave, eventually accelerating
it to arbitrarily high energy \cite{gunnostriker71}. On a limit cycle,
this tendency must be opposed by a counter-propagating wave. Clearly,
the stronger the radiation reaction, the higher must be the amplitude
of the counter-propagating wave. Thus, for fixed $\lambda$, one
expects limit cycles only for $\rr$ less than some maximum value. 
As $\lambda\rightarrow1$, i.e., in the standing wave, 
$\rrmax\rightarrow\infty$, and the limit
cycles approach $\pi/2$ and $\pi$, which are the antinodes of $\bm{B}$ and
$\bm{E}$, and nodes of $\bm{E}$ and $\bm{B}$, respectively. 
At fixed $\lambda$, figure~\ref{xpoints} shows
that the limit cycles are located at these same points when
$\rr\rightarrow0$. In this limit, particles extract no energy from
the fields, implying that their velocity is perpendicular to the
electric field. Although the fields do not have nodes when $\lambda\ne 1$, they are 
perpendicular to each other at $x=\pi/2$ and $x=\pi$, as 
can be seen from Eq~(\ref{edotb}).

\begin{figure}
\includegraphics{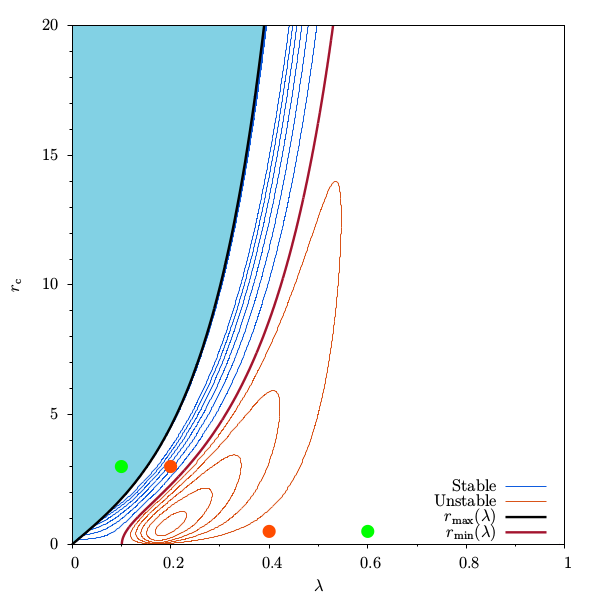}
\caption{\label{contour}
Contour plot of the largest real part of the eigenvalues 
of the Jacobian (\ref{jacobian}), evaluated on the limit cycle at the point $x=x_B$.
Contours are equally spaced, between $-0.02$ and $-0.12$ in the stable region
(blue), and between $0.01$ and $0.06$ in the unstable region (orange). 
Also plotted is the line of marginal stability $\rrmin\left(\lambda\right)$ and the 
line $\rrmax\left(\lambda\right)$,  on which the locations of the 
two limit cycles merge. In the shaded region, where 
$\rr>\rrmax$, no limit cycles are found. Red points mark the parameters used
for the stroboscopic maps in figures~\ref{attractor1fig} and \ref{attractor2fig} and
the histograms in Fig~\ref{histogram12fig}, green points those for the 
histograms in Fig~\ref{histogram34fig}.
 }
\end{figure}

The importance of these limit cycles depends crucially on their
stability. A linear analysis is straightforward to perform, because
the Jacobian $J$ of the system \dynamicalsystem\ is independent of time
when evaluated on a limit cycle, where $\pphi=t+\textrm{constant}$.
After some calculation, one finds that on such a cycle
\begin{align}
J&=
\left(
\begin{array}{cccc}
\partial \dot{x}/\partial x&
\partial \dot{x}/\partial u_x&
\partial \dot{x}/\partial \pr&
\partial \dot{x}/\partial \pphi\\
&&&\\
\partial \dot{u_x}/\partial x&
\dots&
\dots&
\dots\\
&&&\\
\partial \dot{\pr}/\partial x&
\dots&
\dots&
\dots\\
&&&\\
\partial \dot{\pphi}/\partial x&
\dots&
\dots&
\dots\\
\end{array}
\right)
\nonumber\\
& =
\left(
\begin{array}{cccc}
0&
\frac{1}{\hat{\gamma}}&
0&
0\\
&&&\\
\frac{\lambda\cos 2x-\prhat^2}{\hat{\gamma}}&
-\frac{\rr\etahat^2}{\hat{\gamma}}&
-\frac{\lambda\sin 2x}{2\prhat\hat{\gamma}}&
0\\
&&&\\
-\frac{\rr\lambda\etahat^2\sin 2x}{2\prhat\hat{\gamma}}&
0&
-\frac{\rr\etahat^2}{\hat{\gamma}}&
\prhat\\
&&&\\
-\frac{\lambda\hat{\gamma}^2\sin 2x}{\etahat^2}&
-\frac{\left(1-\lambda^2\right)\hat{\gamma}}{2\etahat^2}&
-\frac{\lambda^2\sin^22\theta_0}{2\etahat^2\prhat} -\frac{1}{\prhat}&
\frac{\left[2\lambda\rr\etahat^2\sin 2\theta_0 - \hat{\gamma}\left(1+\lambda^2+2\lambda\cos 2x\right)+2\etahat^2/\hat{\gamma}\right]\prhat^2}
{2\rr\etahat^4}\\
&&&-\frac{\rr\etahat^2}{\hat{\gamma}}\\
\end{array}
\right)
\label{jacobian}
\end{align}
Thus, the limit
cycle is stable when the real part of each eigenvalue 
of (\ref{jacobian}) is negative, and
unstable when the real part of any eigenvalue is positive.  

As $\rr\rightarrow0$, with $\lambda$ held constant, 
one finds that  the four 
eigenvalues $s_E$ of the Jacobian of the limit cycle
at $x=x_E$ tend to the values
\begin{align}
s_E
&\rightarrow\pm\imath\left(1+\lambda\right)/2,\ \pm\sqrt{\lambda}\,,
\end{align}
showing that these cycles are unstable.
On the other hand, on the limit cycle at $x=x_B$, they
tend to purely imaginary values:
\begin{align}
s_B
&\rightarrow\pm\imath\left(1-\lambda\right)/2,\ \pm\imath\sqrt{\lambda}\,,
\end{align}
indicating marginal stability on the $\rr=0$ axis. 
The lowest order term in $\rr$ in an expansion of the real parts 
of these eigenvalues gives
\begin{align}
\textrm{Re}\left(s_B\right)&=-\rr\frac{
\left(1-\lambda\right)^4\left(1-10\lambda+\lambda^2\right)}
{16\left(1-6\lambda+\lambda^2\right)}+\textrm{O}\left(\rr^3\right)
\label{approxeigb1}
\\
\noalign{\hbox{for the first pair, and}}
\textrm{Re}\left(s_B\right)&=-\rr\frac{
\left(1-\lambda\right)^6}
{16\left(1-6\lambda+\lambda^2\right)}+\textrm{O}\left(\rr^3\right)
\label{approxeigb2}
\end{align}
for the second. 
Thus, for sufficiently small $\rr$, the limit cycle is stable 
for $\lambda<5-\sqrt{24}\approx 0.1$, and unstable otherwise.
For arbitrary $\rr$, an
explicit expression for the eigenvalues 
in closed form is cumbersome,
but easily found, using, for example, {\em Mathematica}.

The properties of the limit cycles are shown in
figure~\ref{stationary_sols} for three illustrative values of
$\rr$. Note that the orbits at $x=x_E$ (shown in red) are unstable for
all values of $\lambda$, whereas there exists a range of $\lambda$ in
which the orbits at $x=x_B$ (shown in blue) are stable.  This is shown
in more detail in the contour plot of figure~\ref{contour}. For each
value of $\lambda$, limit cycles exist only for $\rr<\rrmax$. For
$\rrmin<\rr<\rrmax$, the cycles at $x=x_B$ are stable, whereas for
$\rr<\rrmin$, they are unstable. Note, in particular, that for
$\lambda<0.1$ the limit cycle at $x_B$ is stable for all $\rr$, as
suggested by the approximate solutions given in
equations~(\ref{approxeigb1}) and (\ref{approxeigb2}).

\begin{figure}
\includegraphics{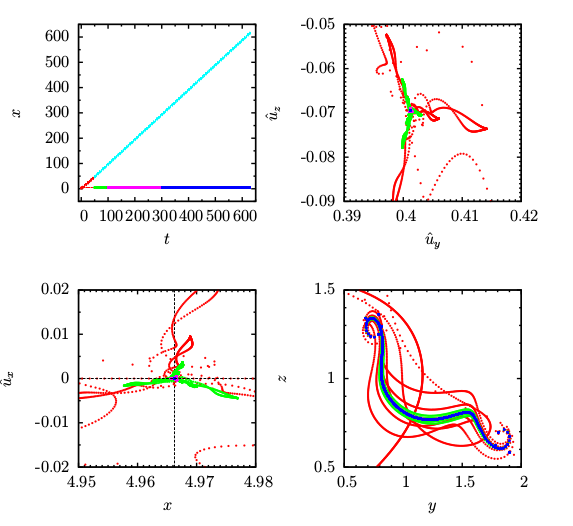}
\caption{\label{attractor1fig} Stroboscopic plots of 5000 trajectories
  integrated for 100 wave periods using parameters $\lambda=0.2$,
  $\rr=3$ that permit a stable limit cycle according to
  figure~\ref{contour}.  The upper left panel shows a clear separation
  into trapped and untrapped particles after a few wave periods. For $t<50$, points are shown
  in red.  For $t>50$, points on untrapped trajectories are plotted in
  light-blue, whereas those on trapped trajectories are plotted in
  colours that change according to the elapsed time, as shown in the
  upper left panel.  In the lower left panel, the $x$--$\hat{u}_x$
  plane is plotted, with the position of the stable limit cycle
  indicated by the intersection of the dashed lines. The right-hand
  panels show the phase space components transverse to the propagation
  direction. }
\end{figure}

\begin{figure}
\includegraphics{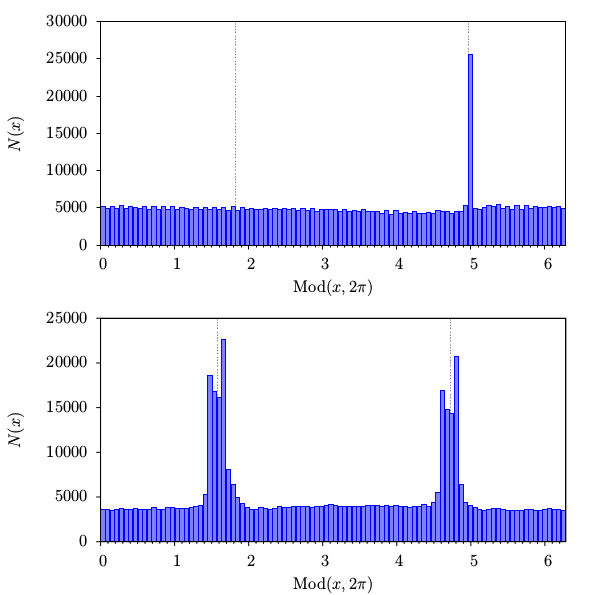}
\caption{\label{histogram12fig}
Histograms of the binned $x$ coordinates.
The upper panel shows the stable case 
corresponding to figure~\ref{attractor1fig} ($\lambda=0.2$, $\rr=3$), 
and the lower panel the
unstable case corresponding to figure~\ref{attractor2fig} 
($\lambda=0.4$, $\rr=.5$). 
The locations of the limit cycles at $x=x_B$ are 
indicated by the dashed lines.
 }
\end{figure}

\section{Stroboscopic maps}
\label{numericalresults}
A linear stability analysis can give only a rough indication of the
relevance of a limit cycle in a physical situation. For this reason,
we have performed numerical integration of the driven dynamical system
\dynamicalsystem\ and constructed stroboscopic maps giving the
locations of the orbits in phase space at times $\hat{t}=2\pi n$,
where $n$ is a positive integer. Results are presented for the choices
of $\lambda$ and $\rr$ marked in figure~\ref{contour} by red dots. For
parameters corresponding to the green dots in this figure we present
only histograms showing the cumulative number of trajectories at a
given value of $\textrm{mod}\left(x,2\pi\right)$.

Figure~\ref{attractor1fig} shows results for $\lambda=0.2$, $\rr=3$,
which lies in the stable region of Fig~\ref{contour}. Trajectories of
5000 positrons are integrated from $t=0$ to $t=200\pi$, initiated at
equally spaced points in $x$ between $0$ and $2\pi$, with
$\hat{u}_x=0$, $\hat{\gamma}=1$ (i.e., $\gamma=a_0$) and values of
$\prhat$ and $\pphi$ corresponding to
$\hat{u}_y=\hat{u}_z=1/\sqrt{2}$. Points depict the location in phase
space at times $t=2n\pi$ for $n=0,1,\dots 200$.  The top left panel
shows the $t$-$x$ plane, with various colour codes that serve to
identify the corresponding points in the other three panels. For
$0<t<50$, points are shown in red.  After only a few wave periods, the
trajectories separate into two groups: those that are picked up by the
primary wave and acquire a high momentum in the positive $x$-direction
(coloured light blue) and those that become trapped close to a limit
cycle near the origin of the $x$-axis. The latter group is subdivided,
as shown, according to the elapsed time.  The remaining panels
concentrate on the trapped particles. The top right panel plots the
components of the four-momentum (normalized to $a_0mc$) in the
transverse ($y$-$z$) plane. The trajectories converge rapidly to a
point, indicating periodic motion. The bottom left panel shows a small
part of the $x$-$\hat{u}_x$ plane, close to the location of the limit
cycle at $x=x_B$ found in section~\ref{analyticalresults}, which is
marked by the vertical dashed line. This panel illustrates rapid
convergence to the limit cycle: for $50<t<100$ (color-coded green),
the trapped particles lie with $\pm0.04$ of the computed position of
the limit cycle, and continue to converge even more tightly at later
times. The lower right panel shows position in the transverse
plane. Here the trajectories converge to a line rather than a point,
indicating the presence of a slow drift of the orbits in this plane
that is not strictly periodic at the fundamental frequency.

\begin{figure}
\includegraphics{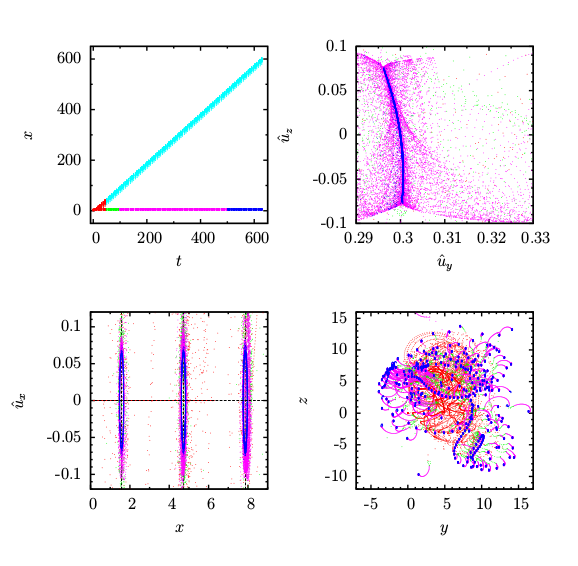}
\caption{\label{attractor2fig} Stroboscopic plots of trajectories
  integrated for 100 wave periods for $\lambda=0.4$ and $\rr=0.5$ ---
  values that do not permit a stable limit cycle according to
  figure~\ref{contour}.  The unstable limit cycles at $x=x_B$,
  $\hat{u}_x=0$ are located at the intersections of the dashed lines
  in the lower left panel.  }
\end{figure}

The concentration of particles around the limit cycle is also clearly
seen in the histogram presented in the upper panel of
figure~\ref{histogram12fig}.  Here, the $x$ values measured at times
$2\pi n$ are mapped onto the interval $[0,2\pi]$ and accumulated in 100 bins.
From the total of 5000 computed trajectories, this figure shows that
approximately 250 are trapped at the location of a stable limit
cycle.

\begin{figure}
\includegraphics{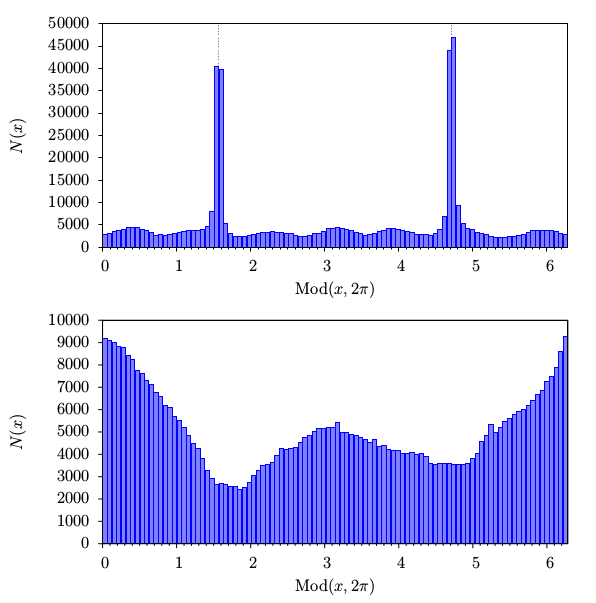}
\caption{\label{histogram34fig} Histograms of the binned $x$
  coordinates for the parameters $\lambda=0.6$, $\rr=0.5$, in the
  unstable region of figure~\ref{contour} (upper panel) and for
  $\lambda=0.1$, $\rr=3$, in the region of figure~\ref{contour} in which
  no limit cycles are found (lower panel). }
\end{figure}
For parameters $\lambda=0.4$, $\rr=0.5$, which lies in the unstable
region of figure~\ref{contour}, the stroboscopic plots are shown in
Fig~\ref{attractor2fig} and the corresponding histogram is the lower
panel of figure~\ref{histogram12fig}. The trajectories no longer
converge exactly onto the limit cycle, but, nevertheless, remain
trapped in its immediate vicinity. The component of the particle
momentum along the wave propagation direction oscillates with an
amplitude of roughly $0.05a_0mc$, and the projection of the trajectory
onto the transverse plane reveals a complex drifting pattern,
superposed on almost circular motion. The lower panel of
Fig~\ref{histogram12fig} shows that particles are trapped in a range
of $x$ that is approximately $10\%$ of one wavelength. Roughly $1500$
of the $5000$ trajectories are trapped --- significantly more than in
the linearly stable case.

Histograms of position are shown in Fig~\ref{histogram34fig} for
parameters corresponding to the green points in figure~\ref{contour}.
As illustrated in the upper panel, strong trapping is a prominent
feature wherever a limit cycle at $x=x_B$ exists, independent of
whether or not it is stable.  For the chosen parameters
($\lambda=0.6$, $\rr=0.5$), approximately 1800 of the 5000
trajectories are confined close to the unstable limit cycle. However,
as can be seen from figure~\ref{stationary_sols}, the limit cycle in
this case has values of $\hat{\gamma}$ and $\etahat$ that are well
approximated by equations~(\ref{gammaapprox}) and (\ref{etaapprox}) and are, therefore, 
substantially smaller than those encountered for the same $\rr$ in the stable range of $\lambda$. 
On the other hand, trapping is absent when no limit cycle
can be found, as shown in the lower panel of figure~\ref{histogram34fig}
for $\lambda=0.1$ and $\rr=3$. In this parameter regime, the radiation
reaction force associated with the stronger wave (wave~1) 
is too strong to be balanced by the weaker wave and accelerates particles to high momentum
in the positive $x$ direction.
  
\section{Discussion}
\label{discussion}

In the absence of radiation reaction, the classical motion of a
charged particle moving in a circularly polarized, monochromatic plane
wave is a combination of periodic motion and a constant drift
speed. Radiation reaction destroys this pattern, causing the particle
to be swept along in the direction of propagation of the wave and
accelerated to arbitrarily high energy. However, the results of the
previous sections show that the presence of a weaker,
counter-propagating, secondary wave can stabilize the motion. In the ZMF, in
which the secondary wave has the same frequency as the
primary, a substantial number of particles accumulate in a plane that
is fixed in configuration space --- they are {\em trapped} at
precisely determined locations on the axis of propagation of the waves
(at $x=x_B$ in the notation of section~\ref{analyticalresults}) and
execute almost circular motion in the plane transverse to this
direction. On these orbits, the particle velocity is always parallel to the 
local magnetic field, as in the case of the self-consistent structures
analyzed in \cite{kirkbellridgers13}.

For this to be possible, the secondary wave must have opposite
helicity to that of the primary. Furthermore, its amplitude must
exceed a certain threshold, which depends on the classical radiation
reaction parameter $\rr$, and can be read off from
figure~\ref{contour}. In terms of the intensity
$I_{24}\times10^{24}\,\wsqcm$ and wavelength
$\lambdamicron\,\mu\textrm{m}$ of the primary wave,
\begin{align}
\rr&=21\times I_{24}^{3/2}\lambdamicron^2\,.
\end{align}
Immediately above the threshold, the limit cycle underlying the
attractor is linearly stable, and particle orbits converge onto it
asymptotically. Linear stability is lost at a point higher above 
threshold, but particles still accumulate very close to the
limit cycle, performing chaotic motion on a strange
  attractor. As the intensity of the secondary wave
approaches that of the primary, a larger fraction of the trajectories sampled
in section~\ref{numericalresults} become trapped, and the growth rate of the linear
instability weakens.  Particles moving on
this limit cycle have orbits that are well-approximated by
equations~(\ref{xbapprox})--(\ref{etaapprox}), unless $\lambda$ is
 very close to or below threshold. The corresponding values of 
the 
  Lorentz factor and QED parameter $\eta$ are
\begin{align}
\gamma&\approx605\times\left(1-\lambda\right) I_{24}^{1/2}\lambdamicron
\\
\eta&=0.89\times \left(1-\lambda\right)^2 I_{24}\lambdamicron\,.
\end{align}
Photons are radiated incoherently by particles moving on the attractor, and 
have a synchrotron-like spectrum, 
which peaks at the energy
\begin{align}
E_\gamma&= 0.65\times \rr\hat{\gamma}^3 \left(mc^2/\alpha_{\textrm{f}}\right)
\\
&\approx120\times\left(1-\lambda\right)^3 I_{24}^{3/2}\lambdamicron^2\,
\textrm{MeV,}
\end{align}
where $\alpha_{\textrm{f}}$ is the fine-structure constant. Thus,
although more trajectories are trapped for larger $\lambda$, this
region results in lower values of $\gamma$ and $\eta$, so that the
optimal value for creating hard gamma-rays and/or observing QED
effects lies close to the threshold.

As a potential application, consider a set-up similar to that adopted
in \cite{liuetal15}, where relativistic, resonant phase-locking in a
single, circularly polarized wave has been proposed as an attractive
mechanism for the production of gamma-rays and relativistic electron
beams. For a primary wave of intensity $I_{24}=0.08$ and
$\lambdamicron=1$ one finds $\rr=0.5$. Thus, introducing a
counter-propagating circularly polarized pulse of opposite helicity,
and with an intensity of roughly 16\% of the primary wave would
reproduce the conditions investigated in figure~\ref{attractor2fig} and
the upper panel of figure~\ref{histogram34fig}, leading to the
emission of gamma-rays of energy roughly $2~\textrm{MeV}$, highly
collimated into a fan beam in the plane perpendicular to the direction
of wave-propagation. For a primary wave with $I_{24}=0.2$ and
$\lambdamicron=1$, a secondary wave with $4\%$ of
this intensity produces an attractor on which $\eta\approx0.1$, which lies 
within the range expected to lead to a non-linear pair cascade
\cite{bellkirk08}.

In additional to the limit cycle that underlies the attractor, another
exists at $x=x_E$. In the limit of a standing wave ($\lambda=1$), this
is the orbit studied by Bell \& Kirk \cite{bellkirk08}, who used it to
analyze the possibility of generating a non-linear cascade of
electron-positron pairs. On it, the values of $\hat{\gamma}$ and
$\etahat$ are much larger than those found on the cycle at
$x=x_B$. However, figure~\ref{stationary_sols} shows that this orbit
is linearly unstable, with a growth rate that is of the same order as
the wave frequency, and substantially exceeds that of the unstable
modes at $x=x_B$. Furthermore, the results of
section~\ref{numericalresults} show no indication of the existence of
an attractor related to this limit cycle, in agreement with subsequent
work on cascades \cite{kirkbellarka09}.

The lack of an attractor with a strong electric field in a circularly
polarized standing wave has led to the suggestion that linear
polarization might provide a more effective means of realizing
non-linear pair cascades \cite{jirkaetal16}.  Attractors of the kind
described above, in which the orbit lies in the plane transverse to
the propagation vector, can be found in this case only if the
polarization vectors of the two waves are precisely aligned, and the
wave intensities are exactly equal. Standing waves in this
configuration have been investigated in \cite{gonoskovetal14}, where
trapping was found not only on an attractor at the analogue of
$x=x_B$, but also, for very high intensities, at the analogue of
$x=x_E$. This phenomenon has no counterpart in the circularly
polarized case discussed above, where the limit cycles at $x_E$ are
strongly unstable at all intensities. However, the limit cycle orbit
in linear polarization is rectilinear. Consequently, the radiation
mechanism --- {\em linear acceleration emission}
\cite{melroserafatluo09,revillekirk10} --- differs significantly from
synchrotron radiation, and equation~(\ref{rrterms}) is not an adequate
approximation to the radiation reaction force \cite{bulanovetal11},
which becomes a very sensitive function of the longitudinal momentum
$u_x$. In addition to modifying the stability properties, this effect
also leads to a relatively low value of the QED parameter $\eta$,
making it more difficult for trapped particles to enter the regime of
strong field QED.

The results presented above are subject to several limitations.
Firstly, they are based on an analysis of the equations of motion in
classical electrodynamics, assuming highly relativistic motion, and
using an approximate formulation of the radiation reaction force due
to \cite{landaulifshitz75}. This requires that the QED effects be
minor, i.e., $\eta\ll1$. For improved accuracy, the classical
radiation reaction term can be modified by the function
$g\left(\eta\right)$ as described by \cite{kirkbellarka09}. This
correction is straightforward to implement, but introduces an
additional parameter, and does not produce any qualitative
changes. For $\eta>0.1$, the influence of {\em straggling}
\cite{duclousetal11,neitzdipiazza13,blackburnetal14} becomes
important, and the structure of the attractors is expected to change
\cite{jirkaetal16}, although it is unlikely that they disappear
entirely. Secondly, a plane-wave approximation is used. The lower
right panels in figures~\ref{attractor1fig} and \ref{attractor2fig}
show that the attractors drift around in the transverse plane over a
region with a typical extension of a few wavelengths. In a realistic
situation, for example, a tightly focused pulse, transverse structure
in the fields might limit the time for which trajectories remain
trapped.  Thirdly, the waves are assumed to be circularly polarized
and monochromatic.  Relaxing this assumption makes an analytical
treatment difficult, and it is not clear whether or not the attractors
described above are generic features.  Fortunately, none of these
limitations applies to numerical simulations in which strong-field QED
effects have been implemented (e.g., \cite{ridgersetal14}), so that
their importance under more realistic condition can readily be
checked.

\section{Conclusions}
\label{conclusions}

The class of electron limit cycles identified in
section~\ref{analyticalresults} underlies the phenomenon of radiative
trapping on attractors. A numerical study
(section~\ref{numericalresults}) shows that they have a simple
structure where the corresponding limit cycle is stable, but that they
also extend into the unstable region, where they become chaotic.  This
enables the electron dynamics in counter-propagating, circularly
polarized laser pulses to be controlled by adjusting the relative
intensity $\lambda$ of the pulses, according to figure~\ref{contour}.  In
intense optical pulses with $I\sim 10^{23}\,\wsqcm$, the radiation reaction
parameter $\rr\sim1$, and the regime of strong-field QED is best
realized using a secondary probe beam with an intensity equal to a few
percent of the primary.

These results are based on the equations of motion of classical
electrodynamics, using a relativistic approximation to the radiation
reaction term as formulated by Landau \& Lifshitz
\cite{landaulifshitz75}. To test their applicability to realistic,
10~PW pulses, a full numerical treatment including quantum effects is
required.

\ack
I thank Antonino di~Piazza, Tony Bell and Chris Ridgers for helpful discussions. 

\section*{References}
\bibliographystyle{unsrt}
\bibliography{references}

\begin{thebibliography}{10}

\bibitem{ELI}
{Extreme Light Infrastructure}.
\newblock \url{http://www.eli-beams.eu}.
\newblock Accessed: 2016-03-25.

\bibitem{XCELS}
{Exawatt Centre for Extreme Light Studies}.
\newblock \url{http://www.xcels.iapras.ru}.
\newblock Accessed: 2016-03-25.

\bibitem{HiPER}
{European High Power Laser Energy Research Facility}.
\newblock \url{http://www.hiper-laser.org}.
\newblock Accessed: 2016-03-25.

\bibitem{dipiazzaetal12}
A.~{Di Piazza}, C.~{M{\"u}ller}, K.~Z. {Hatsagortsyan}, and C.~H. {Keitel}.
\newblock {Extremely high-intensity laser interactions with fundamental quantum
  systems}.
\newblock {\em Reviews of Modern Physics}, 84:1177--1228, July 2012.

\bibitem{dipiazza08}
A.~{di~Piazza}.
\newblock {Exact solution of the Landau-Lifshitz equation}.
\newblock {\em Lett.\ Math.\ Phys.}, 83:305--313, 2008.

\bibitem{mendoncadoveil82}
J.~T. {Mendonca} and F.~{Doveil}.
\newblock {Stochasticity in plasmas with electromagnetic waves}.
\newblock {\em Journal of Plasma Physics}, 28:485--493, December 1982.

\bibitem{bauermulsersteeb95}
D.~{Bauer}, P.~{Mulser}, and W.-H. {Steeb}.
\newblock {Relativistic ponderomotive force, Uphill acceleration, and
  transition to chaos}.
\newblock {\em Physical Review Letters}, 75:4622--4625, December 1995.

\bibitem{lehmannspatschek12}
G.~{Lehmann} and K.~H. {Spatschek}.
\newblock {Phase-space contraction and attractors for ultrarelativistic
  electrons}.
\newblock {\em \pre}, 85(5):056412, May 2012.

\bibitem{zhidkovetal02}
A.~{Zhidkov}, J.~{Koga}, A.~{Sasaki}, and M.~{Uesaka}.
\newblock {Radiation Damping Effects on the Interaction of Ultraintense Laser
  Pulses with an Overdense Plasma}.
\newblock {\em Physical Review Letters}, 88(18):185002, May 2002.

\bibitem{naumovaetal09}
N.~{Naumova}, T.~{Schlegel}, V.T. {Tikhonchuk}, C.~{Labaune}, I.V. {Sokolov},
  and G.~{Mourou}.
\newblock Ponderomotive ion acceleration in dense plasmas at super-high laser
  intensities.
\newblock {\em The European Physical Journal D}, 55(2):393--398, 2009.

\bibitem{tamburinietal10}
M.~{Tamburini}, F.~{Pegoraro}, A.~{Di Piazza}, C.H. {Keitel}, and A.~{Macchi}.
\newblock Radiation reaction effects on radiation pressure acceleration.
\newblock {\em New Journal of Physics}, 12(12):123005, 2010.

\bibitem{chenetal11}
M.~{Chen}, A.~{Pukhov}, T-P. {Yu}, and Z-M. {Sheng}.
\newblock Radiation reaction effects on ion acceleration in laser foil
  interaction.
\newblock {\em Plasma Physics and Controlled Fusion}, 53(1):014004, 2011.

\bibitem{nakamuraetal12}
T.~{Nakamura}, J.~{Koga}, T.~{Esirkepov}, M.~{Kando}, G.~{Korn}, and S.V.
  {Bulanov}.
\newblock High-power gamma-ray flash generation in ultraintense laser-plasma
  interactions.
\newblock {\em Physical Review Letters}, 108(19), 2012.

\bibitem{ridgersetal14}
C.~P. {Ridgers}, J.~G. {Kirk}, R.~{Duclous}, T.~G. {Blackburn}, C.~S. {Brady},
  K.~{Bennett}, T.~D. {Arber}, and A.~R. {Bell}.
\newblock {Modelling gamma-ray photon emission and pair production in
  high-intensity laser-matter interactions}.
\newblock {\em Journal of Computational Physics}, 260:273--285, March 2014.

\bibitem{bulanovetal15}
S.~V. {Bulanov}, T.~Z. {Esirkepov}, M.~{Kando}, J.~{Koga}, K.~{Kondo}, and
  G.~{Korn}.
\newblock {On the problems of relativistic laboratory astrophysics and
  fundamental physics with super powerful lasers}.
\newblock {\em Plasma Physics Reports}, 41:1--51, January 2015.

\bibitem{gonoskovetal14}
A.~{Gonoskov}, A.~{Bashinov}, I.~{Gonoskov}, C.~{Harvey}, A.~{Ilderton},
  A.~{Kim}, M.~{Marklund}, G.~{Mourou}, and A.~{Sergeev}.
\newblock {Anomalous Radiative Trapping in Laser Fields of Extreme Intensity}.
\newblock {\em Physical Review Letters}, 113(1):014801, July 2014.

\bibitem{jietal14}
L.~L. {Ji}, A.~{Pukhov}, I.~Y. {Kostyukov}, B.~F. {Shen}, and K.~{Akli}.
\newblock {Radiation-Reaction Trapping of Electrons in Extreme Laser Fields}.
\newblock {\em Physical Review Letters}, 112(14):145003, April 2014.

\bibitem{esirkepovetal15}
T.~Z. {Esirkepov}, S.~S. {Bulanov}, J.~K. {Koga}, M.~{Kando}, K.~{Kondo}, N.~N.
  {Rosanov}, G.~{Korn}, and S.~V. {Bulanov}.
\newblock {Attractors and chaos of electron dynamics in electromagnetic
  standing waves}.
\newblock {\em Physics Letters A}, 379:2044--2054, September 2015.

\bibitem{jirkaetal16}
M.~Jirka, O.~Klimo, S.~V. Bulanov, T.~Zh. Esirkepov, E.~Gelfer, S.~S. Bulanov,
  S.~Weber, and G.~Korn.
\newblock Electron dynamics and $\ensuremath{\gamma}$ and
  ${e}^{\ensuremath{-}}{e}^{+}$ production by colliding laser pulses.
\newblock {\em Phys. Rev. E}, 93:023207, Feb 2016.

\bibitem{kirkbellarka09}
J.~G. {Kirk}, A.~R. {Bell}, and I.~{Arka}.
\newblock {Pair production in counter-propagating laser beams}.
\newblock {\em Plasma Physics and Controlled Fusion}, 51(8):085008, August
  2009.

\bibitem{landaulifshitz75}
L.~D. {Landau} and E.~M. {Lifshitz}.
\newblock {\em {The Classical Theory of Fields}}.
\newblock Course of theoretical physics - Pergamon International Library of
  Science, Technology, Engineering and Social Studies, Oxford: Pergamon Press,
  1975, 4th rev.engl.ed., 1975.

\bibitem{kirkbellridgers13}
J.~G. {Kirk}, A.~R. {Bell}, and C.~P. {Ridgers}.
\newblock {Pair plasma cushions in the hole-boring scenario}.
\newblock {\em Plasma Physics and Controlled Fusion}, 55(9):095016, September
  2013.

\bibitem{gunnostriker71}
J.~E. {Gunn} and J.~P. {Ostriker}.
\newblock {On the Motion and Radiation of Charged Particles in Strong
  Electromagnetic Waves. I. Motion in Plane and Spherical Waves}.
\newblock {\em \apj}, 165:523, May 1971.

\bibitem{liuetal15}
B.~{Liu}, R.~H. {Hu}, H.~Y. {Wang}, D.~{Wu}, J.~{Liu}, C.~E. {Chen},
  J.~{Meyer-ter-Vehn}, X.~Q. {Yan}, and X.~T. {He}.
\newblock {Quasimonoenergetic electron beam and brilliant gamma-ray radiation
  generated from near critical density plasma due to relativistic resonant
  phase locking}.
\newblock {\em Physics of Plasmas}, 22(8):080704, August 2015.

\bibitem{bellkirk08}
A.~R. {Bell} and J.~G. {Kirk}.
\newblock {Possibility of Prolific Pair Production with High-Power Lasers}.
\newblock {\em Physical Review Letters}, 101(20):200403--+, November 2008.

\bibitem{melroserafatluo09}
D.~B. {Melrose}, M.~Z. {Rafat}, and Q.~{Luo}.
\newblock {Linear Acceleration Emission. I. Motion in a Large-Amplitude
  Electrostatic Wave}.
\newblock {\em \apj}, 698:115--123, June 2009.

\bibitem{revillekirk10}
B.~{Reville} and J.~G. {Kirk}.
\newblock {Linear Acceleration Emission in Pulsar Magnetospheres}.
\newblock {\em \apj}, 715:186--193, May 2010.

\bibitem{bulanovetal11}
S.~V. {Bulanov}, T.~Z. {Esirkepov}, M.~{Kando}, J.~K. {Koga}, and S.~S.
  {Bulanov}.
\newblock {Lorentz-Abraham-Dirac versus Landau-Lifshitz radiation friction
  force in the ultrarelativistic electron interaction with electromagnetic wave
  (exact solutions)}.
\newblock {\em \pre}, 84(5):056605, November 2011.

\bibitem{duclousetal11}
R.~{Duclous}, J.~G. {Kirk}, and A.~R. {Bell}.
\newblock {Monte Carlo calculations of pair production in high-intensity
  laser-plasma interactions}.
\newblock {\em Plasma Physics and Controlled Fusion}, 53(1):015009, January
  2011.

\bibitem{neitzdipiazza13}
N.~{Neitz} and A.~{Di Piazza}.
\newblock {Stochasticity Effects in Quantum Radiation Reaction}.
\newblock {\em Physical Review Letters}, 111(5):054802, August 2013.

\bibitem{blackburnetal14}
T.~G. {Blackburn}, C.~P. {Ridgers}, J.~G. {Kirk}, and A.~R. {Bell}.
\newblock {Quantum Radiation Reaction in Laser-Electron-Beam Collisions}.
\newblock {\em Physical Review Letters}, 112(1):015001, January 2014.

\end{thebibliography}
\end{document}